\documentclass[aps,prl,reprint,citeautoscript,superscriptaddress,showpacs]{revtex4-1}

\usepackage{graphicx,epsfig,float}
\usepackage{amssymb} 
\usepackage{amsmath}

\begin{document}
\title{Two phases with different domain wall networks and a reentrant phase transition in bilayer graphene under strain}

\author{Irina V. Lebedeva}
\email{liv\_ira@hotmail.com}
\affiliation{CIC nanoGUNE BRTA,
San Sebasti\'an 20018, Spain}
\author{Andrey M. Popov}
\email{popov-isan@mail.ru}
\affiliation{Institute for Spectroscopy of Russian Academy of Sciences, Troitsk, Moscow 108840, Russia}

\begin{abstract}
The analytical two-chain Frenkel-Kontorova model is used to describe domain wall networks in bilayer graphene upon biaxial stretching of one of the layers. We show that the commensurate-incommensurate phase transition leading to formation of a regular triangular domain wall network at the relative biaxial elongation of $3.0\cdot 10^{-3}$  is followed by the transition to another incommensurate phase with a striped network at the elongation of $3.7\cdot 10^{-3}$. The reentrant transition to the phase with a triangular domain wall network is predicted for the elongation $\sim10^{-2}$.
\end{abstract}
\maketitle

The presence of two layers in bilayer graphene gives rise to such interesing physical phenomena as topological confinement \cite{Martin2008} and superconductivity \cite{Cao2018}. Here we predict the reentrant phase transition in bilayer graphene under biaxial stretching of one of the layers.
The phenomenon of reentrant phase transition means that upon a monotonic change of any thermodynamic parameter (for example, temperature) two (or more) phase transitions occur and the system finally gets into a state which is macroscopically similar to the initial one \cite{Narayanan1994}. Reentrant phase transitions have been observed for a wide set of 3D systems such as liquid mixtures  \cite{Narayanan1994}, liquid crystals \cite{Singh2000}, microemulsions \cite{Sorensen1985},  granular superconductors \cite{Lin1984}, {\it etc}. The possibility of inverse melting of a polimeric system has been also proposed \cite{Greer2000}. As for 2D systems, a reentrant transition has been observed only for the magnetic domain structure of a thin film \cite{Portmann2003}. 

Some years ago the commensurate-incommensurate phase transition in bilayer graphene under uniaxial elongation of one of the layers was predicted \cite{Popov2011} based on the analytical two-chain Frenkel-Kontorova model  \cite{Bichoutskaia2006, Popov2009}. Bilayer graphene has two types of degenerate but topologically inequivalent minima at the AB and BA stackings with half of the atoms of one layer on top of the atoms of the second layer and the other half on top of the centers of the hexagons. The commensurate-incommensurate phase transition under strain corresponds to formation of a stacking dislocation separating commensurate domains with the AB and BA stackings, that is stacking dislocations play the role of domain walls with incommensurate stacking within the walls. 

Since the prediction of the commensurate-incommensurate phase transition \cite{Popov2011}, domain wall networks have been observed in bilayer and few-layer graphene by various methods  \cite{Alden2013,Lin2013,Kisslinger2015,Jiang2016,Huang2018,Yoo2019}. It has been shown that domain walls influence
electronic \cite{Wright2011, Ju2015, Huang2018, Yoo2019, Vaezi2013, Zhang2013, Hattendorf2013, San-Jose2013, San-Jose2014, Benameur2015, Koshino2013, Efimkin2018, Gargiulo2018, Ramires2018, Rickhaus2018, Yin2016}, magnetic \cite{Kisslinger2015, Wijk2015, Rickhaus2018} and optical \cite{Gong2013} properties of bilayer graphene. Thus, studies of domain wall networks are of  interest not only for foundamental physics but also for graphene applications.

\begin{figure}
	\centering
	\includegraphics[width=\columnwidth]{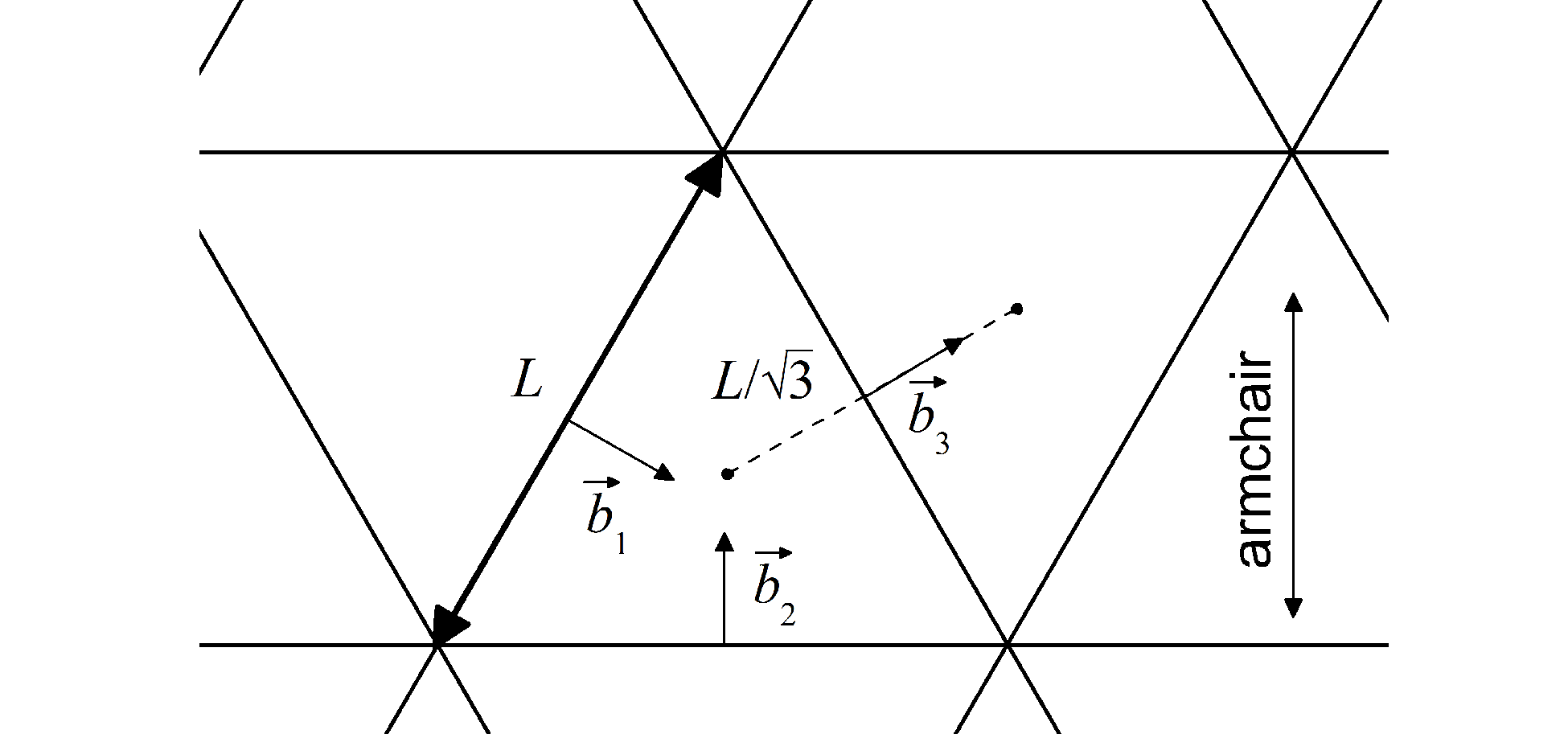}
	\caption{Scheme of a regular triangular network of tensile domain walls in bilayer graphene. The size $L$ of commensurate domains and Burgers vectors $\vec{b}_1$,  $\vec{b}_2$ and $\vec{b}_3$ of the domain walls ($|\vec{b}_1|=|\vec{b}_2|=|\vec{b}_3|=b$) are indicated.} 
	\label{fig:tri}
\end{figure}

\begin{figure}
	\centering
	\includegraphics[width=\columnwidth]{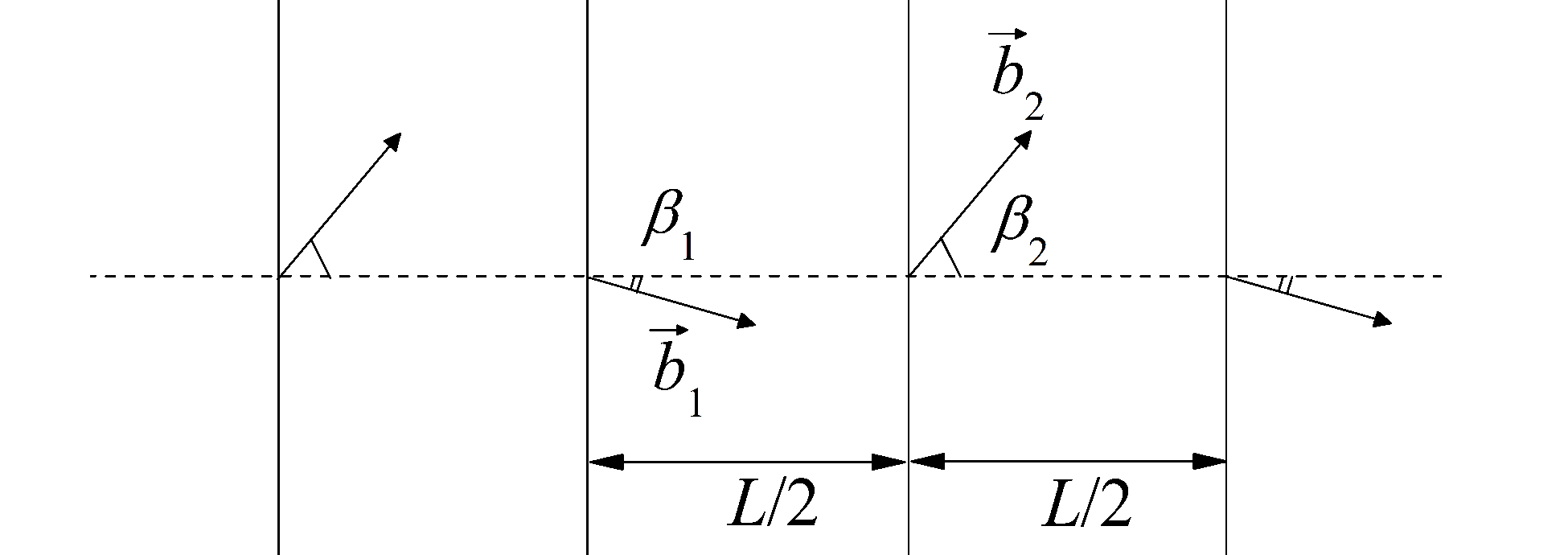}
	\caption{Scheme of a striped domain wall network in bilayer graphene. The size $L/2$ of commensurate domains, Burgers vectors $\vec{b}_1$ and $\vec{b}_2$  of adjacent domain walls ($|\vec{b}_1|=|\vec{b}_2|=b$) and angles $\beta_1$ and $\beta_2$ between the Burgers vectors and normal to the walls are indicated. } 
	\label{fig:striped}
\end{figure}

Recently the two-chain Frenkel-Kontorova model has been extended to take into account dislocation nodes (i.e. crossings of domain walls) and applied to study the commensurate-incommensurate phase transition in bilayer graphene under biaxial elongation of one of the layers \cite{Lebedeva2019, Lebedeva2019a}. A regular triangular domain wall network (Fig. \ref{fig:tri}) was predicted to form as a result of such a transition. However, not only triangular  \cite{Alden2013,Kisslinger2015,Jiang2016,Huang2018,Yoo2019} but also striped networks of parallel domain walls (Fig. \ref{fig:striped}) were observed in the experiments  \cite{Alden2013,Lin2013,Kisslinger2015} where biaxial strain in the bottom layer could be induced by the interaction with the substrate. In some cases both of the network structures were found within the same sample \cite{Alden2013,Kisslinger2015}. Symmetry lowering through formation of a striped network can be energetically favourable under certain conditions because such a network does not involve energetically expensive dislocation nodes. This is the case, for example, for rare gas monolayers adsorbed on graphite \cite{Bak1979}. Here we apply the two-chain Frenkel-Kontorova model to study energetics of striped  domain wall networks in bilayer graphene and to demonstrate that there should be a phase transition between the incommensurate phases with regular triangular and striped domain wall networks (triangular and striped incommensurate phases) under biaxial stretching of one of the graphene layers.

At large elongations, the interlayer interaction is not able to compete with the elastic energy and only weakly perturbs bilayer structure. In this case, moir\'e patterns \cite{Pochet2017,Gargiulo2018,Wijk2015}, in which the size of commensurate domains is comparable to the width of domain walls, are observed. Such superstructures, nevertheless, have the same symmetry as the structures with regular triangular domain wall networks and  belong to the same triangular incommensurate phase. Therefore, the reentrant phase transition to the triangular phase is inevitable at large strains applied. 

The analytical description of the commensurate-incommensurate phase transition in bilayer graphene \cite{Popov2011,Lebedeva2016,Lebedeva2019,Lebedeva2019a} is based on the approximation of the potential energy surface for interlayer interaction  by the first Fourier harmonics \cite{Lebedeva2011,Popov2012,Reguzzoni2012,Lebedeva2010,Lebedeva2011a}: 
\begin{equation} \label{eq_pes}
\begin{split}
V(u_x,u_y) =  &2V_\mathrm{max}\bigg(3/2+\cos\Big(2k_0u_x-2\pi/3\Big) \\
&-2\cos\Big(k_0u_x-\pi/3\Big)\cos\Big(k_0u_y\sqrt{3}\Big)\bigg).
\end{split}
\end{equation} 
Here $V_\mathrm{max}$ is the barrier to relative in-plane motion of the layers, $u_x$ ($u_y$) is the relative displacement in the armchair (zigzag) direction, $k_0$ is determined by the bond length $l$ of graphene as $k_0 = 2\pi/(3l)$ and the energy is given with respect to the AB (BA) stacking. The stacking dislocations in bilayer graphene are partial. The Burgers vectors are equal in magnitude to the bond length of graphene, $b=l$, and aligned along armchair directions. The layers are displaced along the straight minimum energy paths between adjacent AB and BA minima (for example, $u_x=0-l$, $u_y=0$) \cite{Popov2011,Lebedev2015,Lebedeva2016,Lebedeva2019,Lebedeva2019a}. Depending on the angle $\beta$ between the Burgers vector and normal to the domain wall, the character of the wall can change from tensile (for the walls aligned in the zigzag direction with $\beta=0$) to shear (for the walls aligned in the armchair direction with $\beta=\pi/2$).

The formation energy of domain walls per unit length is given by 
\cite{Lebedev2015, Lebedeva2016,Lebedeva2019,Lebedeva2019a}
\begin{equation} \label{eq_DW}
W_\mathrm{D}(\beta) =W_\mathrm{0} f(\beta)=\sqrt{\frac{kl^2 V_\mathrm{max}}{(1-\nu^2)}}\left(\frac{3\sqrt{3}}{\pi}-1\right)f(\beta).
\end{equation}
where $k$ is the elastic constant under uniaxial stress expressed as $k=Yh$ through Young's modulus $Y$ and thickness of graphene layers $h$, $\nu$ is Poisson's ratio and $f(\beta)=(\cos^2 \beta +  \sin^2 \beta \cdot (1-\nu)/2)^{1/2}$ describes the dependence of the elastic constant on the angle $\beta$ between the Burgers vector and normal to domain walls. The characteristic domain wall width (dislocation width) equals
\cite{Lebedev2015, Lebedeva2016,Lebedeva2019,Lebedeva2019a}
\begin{equation} \label{eq_l}
	l_\mathrm{D}(\beta) =l_\mathrm{0} f(\beta)=\frac{l}{2} \sqrt{\frac{k}{V_\mathrm{max}(1-\nu^2)}}f(\beta).
\end{equation}

The accuracy of estimates within the two-chain Frenkel-Kontorova model is mostly limited by the uncertainty in the value of the barrier $V_\mathrm{max}$ to relative sliding of the layers. As discussed in our previous papers \cite{Lebedeva2019, Lebedeva2019a}, the scatter in the available first-principles data on $V_\mathrm{max}$ corresponds to the error of about 40\% in $V_\mathrm{max}$ and 20\% in the energy and width of domain walls. Another source of errors is neglect of out-of-plane buckling \cite{Butz2014,Lin2013}. However, such a buckling is strongly suppressed in supported bilayers \cite{Alden2013,Lin2013,Yankowitz2014}. Using the parameters $l = 1.430$~\AA, $k = 331$~J/m$^2$, $\nu = 0.174$ and $V_\mathrm{max} = 1.61$~meV per atom of one of the layers \cite{Lebedeva2016}, we get the widths of 13.4 nm and 8.6 nm for tensile and shear domain walls, respectively. These values are within the 20\% error bar from the experimental values for supported bilayers \cite{Alden2013,Lin2013,Yankowitz2014} of 11 nm and 6 -- 7 nm for tensile and shear walls, respectively, and close to the results of atomistic \cite{Gargiulo2018} and multiscale \cite{Zhang2018} simulations. The use of the Frenkel-Kontorova model is justified as long as the size of commensurate domains is much greater than the domain wall width.

The energy of the bilayer with a domain wall network with respect to the commensurate system with the same relative biaxial elongation $\epsilon$ of the bottom layer per unit bilayer area, i.e. the formation energy of the domain wall network, can be written as a sum of three terms \cite{Lebedeva2019, Lebedeva2019a}:
\begin{equation} \label{eq_DWtot}
\Delta W_\mathrm{tot} = \Delta W_\mathrm{el} + \Delta W_\mathrm{dw}+ \Delta W_\mathrm{dn},
\end{equation}
where $\Delta W_\mathrm{dw}$ and $\Delta W_\mathrm{dn}$ are the contributions of domain walls and dislocation nodes, respectively,  and $\Delta W_\mathrm{el}$ is the term related to the global change of the elongation of the bottom layer upon formation of domain walls.

Let us first give a brief overview of the results \cite{Lebedeva2019,Lebedeva2019a} for the regular triangular network of domain walls with the side $L$ of commensurate domains. As shown in our previous paper \cite{Lebedeva2019a}, if the upper layer is free, it stays co-aligned with the bottom layer upon biaxial stretching of the latter and the triangular network of tensile domain walls aligned in zigzag directions is formed. Therefore, here we limit our consideration to tensile domain walls  (Fig. \ref{fig:tri}). The area of each commensurate domain is $S=\sqrt{3}L^2/4$ and  the total length of domain walls per domain is $3L/2$. The contribution of domain walls to the formation energy of the regular triangular domain wall network in Eq. (\ref{eq_DWtot}) can thus be calculated as $\Delta W_\mathrm{dw}=3LW_\mathrm{0}/(2S)=2\sqrt{3}W_\mathrm{0}/L$. 

In the case of tensile domain walls, it can be assumed that dislocation nodes have the shape of hexagons with the side $l_0$ and that the graphene layers are uniformly stretched and fully incommensurate inside them
\cite{Lebedeva2019, Lebedeva2019a}. As shown in our previous paper \cite{Lebedeva2019}, these assumptions correspond to the error of 10--20\% in the formation energy, which is comparable to the error coming from the scatter in the available data on $V_\mathrm{max}$ 
\cite{Lebedeva2019, Lebedeva2019a} and, therefore, acceptable for our model. The energy per unit area of a dislocation node equals $V_\mathrm{el}+V_\mathrm{in}$, where $V_\mathrm{in}=3V_\mathrm{max}$ is the average interlayer interaction energy over the potential energy surface (\ref{eq_pes}) and $V_\mathrm{el}$ is the elastic energy coming from the tensile strain of $\pm l/(2l_\mathrm{0})$. There is a half of the dislocation node per commensurate domain, 
i.e. the nodes occupy the fraction $3(l_0/L)^2$ of the bilayer area. From these considerations, the contribution of dislocation nodes can be written as \cite{Lebedeva2019, Lebedeva2019a}
\begin{equation} \label{eq_dn}
\Delta W_\mathrm{dn}= 3k\left(\frac{l}{L}\right)^2\frac{5+2\nu}{4(1-\nu^2)}.
\end{equation}

To estimate the term coming from the global change of the elongation of the bottom layer upon formation of the domain wall network in Eq. (\ref{eq_DWtot}), we consider the relative displacement of the layers in adjacent commensurate domains, which increases by the magnitude of the Burgers vector, $b=l$, along the line connecting the centers of the domains. Since the distance between these centers is $L/\sqrt{3}$, it can be deduced that formation of the regular triangular network is associated with the relative biaxial elongation of $\epsilon_0=\sqrt{3}l/(2L)$ in each of the layers. That is to accomodate the triangular domain wall network,  the elongation of the bottom layer increases by $\epsilon_0$ in comparison with the initial commensurate system. To compare the energies of the bilayer with and without the triangular domain wall network at the same biaxial elongation of the bottom layer, we take into account the following term: 
\begin{equation} \label{eq_Wel}
\begin{split}
 \Delta W_\mathrm{el} &=-\frac{2k}{(1-\nu)}\left(\epsilon^2-\left(\epsilon-\epsilon_0\right)^2\right)\\&=-\frac{2\sqrt{3}k\epsilon}{(1-\nu)}\frac{l}{L}+\frac{3k}{2(1-\nu)}\left(\frac{l}{L}\right)^2.
\end{split}
\end{equation}

Finally, on the basis of Eq. (\ref{eq_DWtot}), we arrive at the expression for the formation energy of the regular triangular domain wall network of the form:
\begin{equation} \label{eq_WAB}
\Delta W_\mathrm{tot}= \frac{Al}{L} +\frac{Bl^2}{L^2},
\end{equation}
where 
\begin{equation} \label{eq_A}
A=-\frac{2\sqrt{3} k\epsilon}{1-\nu}+ \frac{2\sqrt{3}W_\mathrm{0}}{l}
\end{equation}
and
\begin{equation} \label{eq_B}
B=\frac{3k(7+4\nu)}{4(1-\nu^2)}.
\end{equation}

This expression shows that as long as the elongation is small and $A$ is positive, the optimal period $L_0$ of the network determined by conditions $\partial \Delta W_\mathrm{tot}/\partial L=0$ and $\partial^2 \Delta W_\mathrm{tot}/\partial L^2\ge0$ tends to infinity, i.e. the commensurate state is more energetically favourable than the systems with triangular domain wall networks. Once $A$ becomes negative, the optimal period of the domain wall network changes as $L_0 = -2Bl/A$
and the formation energy as $\Delta W_0 = -A^2/(4B)$.

\begin{figure}
	\centering
	\includegraphics[width=\columnwidth]{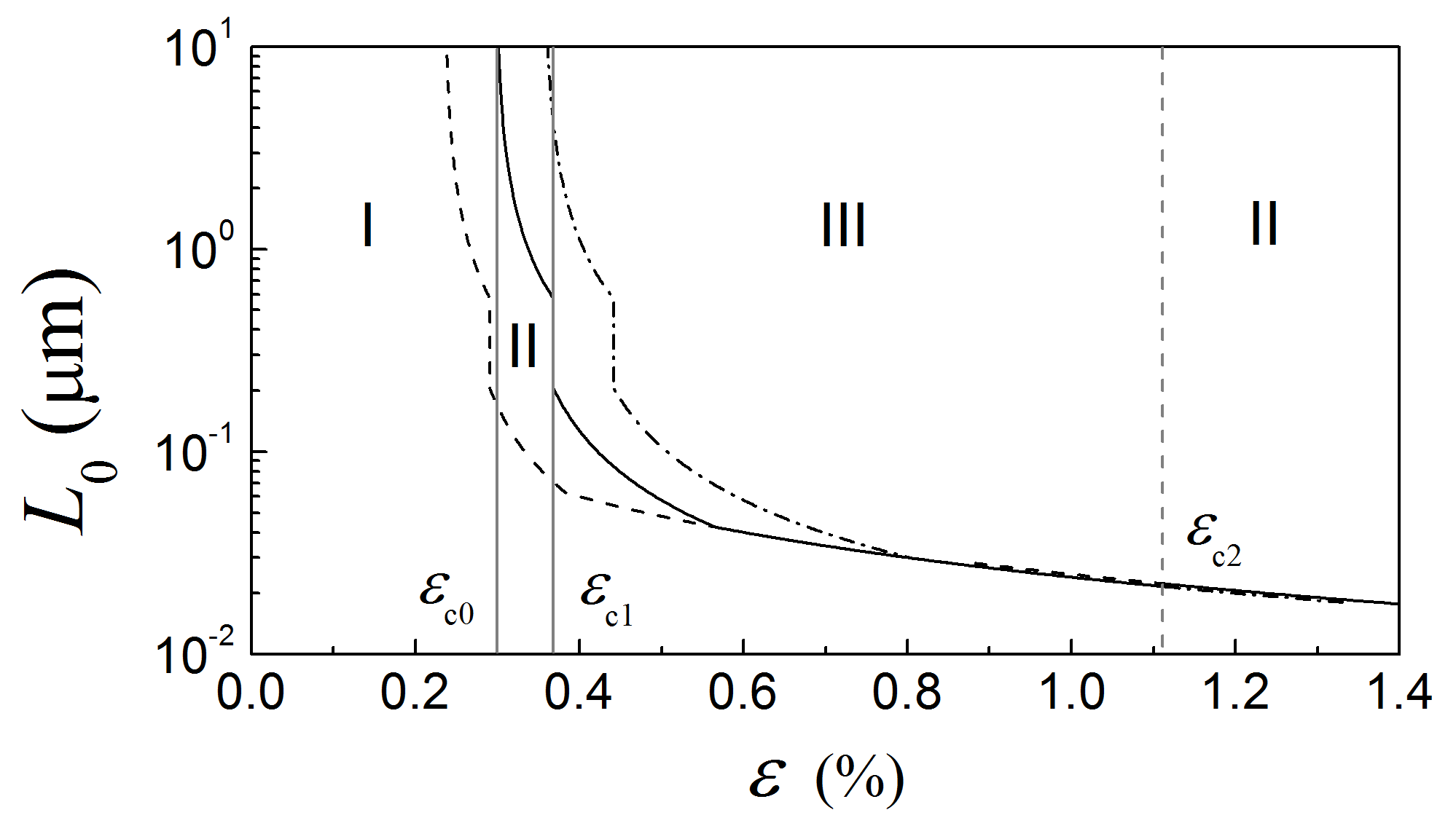}
	\caption{Period $L_0$ of the superstructure of bilayer graphene (in $\mu$m) as a function of relative biaxial elongation $\epsilon$ (in \%) of the bottom layer (solid lines). Different phases are indicated: I - commensurate, II - regular triangular incommensurate, III - striped incommensurate. The dashed (dash-dotted) lines represent the results for the barrier $V_\mathrm{max}$ smaller (greater) by 40\%.} 
	\label{fig:length}
\end{figure}

The transition between the commensurate phase and incommensurate phase with the regular triangular domain wall network happens when $A=0$, i.e. the relative biaxial elongation reaches the critical value $\epsilon_\mathrm{c0}=(1-\nu)W_\mathrm{0}/(kl)=2.97\cdot 10^{-3}$. The expression for the formation energy of the most favourable regular triangular network above the critical elongation  $\epsilon_\mathrm{c}$ can be written as 
\begin{equation} \label{eq_W0}
\Delta W_0  = -\frac{4(1+\nu)}{(1-\nu)(7+4\nu)}k(\epsilon-\epsilon_\mathrm{c0})^2
\end{equation}
and the network period (Fig. \ref{fig:length}) as 
\begin{equation} \label{eq_L0}
L_0 = \frac{\sqrt{3}(7+4\nu)}{4(1+\nu)}\frac{l}{\epsilon-\epsilon_\mathrm{c0}}.
\end{equation} 
As discussed previously \cite{Lebedeva2019,Lebedeva2019a}, such dependences correspond to the second-order phase transition.

Let us now consider the striped incommensurate phase with parallel domain walls  (Fig. \ref{fig:striped}). 
In such a phase, there are no dislocation nodes and $\Delta W_\mathrm{dn}=0$. Since AB and BA domains have to alternate, in bilayer graphene the network should consist of domain walls with alternating directions of the Burgers vector with the angle $\pi/3$ between them \cite{Lebedeva2016}. We assume that the angles between the Burgers vectors and normal to the domain walls for the adjacent walls are $\tilde\beta\pm\pi/6$. As there are two domain walls per network period $L$, the contribution of domain walls for the striped network is given by $\Delta W_\mathrm{dw}(\tilde\beta)= F(\tilde\beta)W_\mathrm{0}/L$, where $F(\tilde\beta)=f(\tilde\beta-\pi/6)+f(\tilde\beta+\pi/6)$. 

The extra elongation is induced by the striped domain wall network in the direction perpendicular to the domain walls and equals $\epsilon_0=(\cos(\tilde\beta-\pi/6)+\cos(\tilde\beta+\pi/6))l/(2L)=\sqrt{3}l\cos{\tilde\beta}/(2L)$. Additionally there is also an extra shear strain $\tau_0=(\sin(\tilde\beta-\pi/6)+\sin(\tilde\beta+\pi/6))l/(2L)=\sqrt{3}l\sin{\tilde\beta}/(2L)$ coming from the relative displacement of the layers along the domain walls. Therefore, the elastic energy term related to the extra elongation and shear strain can be written as
\begin{equation} \label{eq_sWel}
\begin{split}
& \Delta W_\mathrm{el} (\tilde\beta)=-\frac{2k\epsilon\epsilon_0}{(1-\nu)}+\frac{k\epsilon_0^2}{(1-\nu^2)}+\frac{k}{2(1+\nu)}\tau_0^2\\&=-\frac{\sqrt{3}k\epsilon\cos{\tilde\beta}}{1-\nu}\frac{l}{L}+\frac{3k\cos^2{\tilde\beta}}{4(1-\nu^2)}\left(\frac{l}{L}\right)^2+\frac{3k\sin^2{\tilde\beta}}{8(1+\nu)}\left(\frac{l}{L}\right)^2.
\end{split}
\end{equation}

The formation energy of the striped domain wall network can then be presented in the form described by Eq. (\ref{eq_WAB}) with 
\begin{equation} \label{eq_sA}
A^\mathrm{s}(\tilde\beta)=-\frac{\sqrt{3}k\epsilon\cos{\tilde\beta}}{1-\nu}+ \frac{W_\mathrm{0}F(\tilde\beta)}{l}
\end{equation}
and
\begin{equation} \label{eq_sB}
\begin{split}
B^\mathrm{s}(\tilde\beta)=3k\frac{2\cos^2{\tilde\beta}+(1-\nu)\sin^2{\tilde\beta}}{8(1-\nu^2)}.
\end{split}
\end{equation}

The same as for triangular domain wall networks, systems with striped networks become more energetically favourable than the commensurate state when $A^\mathrm{s}(\tilde\beta)=0$. This corresponds to the critical relative biaxial elongation $\epsilon^\mathrm{s}_\mathrm{c}(\tilde\beta)=(1-\nu)W_\mathrm{0}F(\tilde\beta)/(\sqrt{3}kl\cos{\tilde\beta})$.

It is easy to check that $F(\tilde\beta)/\cos{\tilde\beta}$ reaches its minimum for $\tilde\beta=0$. Therefore, the minimal critical relative biaxial elongation  corresponds to $\epsilon^\mathrm{s}_\mathrm{c0}=\epsilon^\mathrm{s}_\mathrm{c}(0)=\epsilon_\mathrm{c0}\sqrt{(7-\nu)/6}=3.17\cdot 10^{-3}$. Since this critical elongation is greater than the value $\epsilon_\mathrm{c0}$ for regular triangular domain wall networks, the latter one indeed corresponds to the commensurate-incommensurate phase transition. 

For each $\tilde\beta$ the relative energy of the bilayer with the optimal striped domain wall network changes above the critical elongation $\epsilon^\mathrm{s}_\mathrm{c}(\tilde\beta)$ as 
\begin{equation} \label{eq_sW0}
\begin{split}
\Delta W_0^\mathrm{s} (\tilde\beta)= -\frac{3k^2(\epsilon-\epsilon^\mathrm{s}_\mathrm{c}(\tilde\beta))^2}{4(1-\nu)^2} \frac{\cos^2{\tilde\beta}}{B^\mathrm{s}(\tilde\beta)}.
\end{split}
\end{equation}
Since $\cos^2{\tilde\beta}/B^\mathrm{s}(\tilde\beta)$ is maximal for $\tilde\beta=0$ (see Eq. (\ref{eq_sB})) and $\epsilon^\mathrm{s}_\mathrm{c}$ is minimal for the same $\tilde\beta$, the most energetically favourable striped network for $\epsilon\ge\epsilon^\mathrm{s}_\mathrm{c0}$ corresponds to $\tilde\beta=0$, i.e. the domain walls with $\beta=\pm\pi/6$ aligned the same armchair direction. For such a domain wall network, the formation energy changes at $\epsilon\ge\epsilon^\mathrm{s}_\mathrm{c0}$  as
\begin{equation} \label{eq_sW0_1}
\Delta W^\mathrm{s}_0 (0) = -\frac{1+\nu}{1-\nu}k(\epsilon-\epsilon^\mathrm{s}_\mathrm{c0} )^2
\end{equation}
and the network period (Fig. \ref{fig:length}) as 
\begin{equation} \label{eq_sL0}
L^\mathrm{s}_0(0)= \frac{\sqrt{3}}{2(1+\nu)}\frac{l}{\epsilon-\epsilon^\mathrm{s}_\mathrm{c0}}.
\end{equation}

From Eqs. (\ref{eq_W0}) and (\ref{eq_sW0_1}), we can estimate that the striped incommensurate phase becomes more energetically favourable than the triangular one at
\begin{equation} \label{eq_c2}
\begin{split}
\epsilon_\mathrm{c1}=\frac{\epsilon^\mathrm{s}_\mathrm{c0}\sqrt{7+4\nu}-2\epsilon_\mathrm{c0}}{\sqrt{7+4\nu}-2}=\epsilon_\mathrm{c0}\frac{\sqrt{\frac{(7+4\nu)(7-\nu)}{6}}-2}{\sqrt{7+4\nu}-2},
\end{split}
\end{equation}
which gives $\epsilon_\mathrm{c1}=1.24\epsilon_\mathrm{c0}=3.68\cdot 10^{-3}$. At this second critical relative biaxial elongation, the size of triangular commensurate domains  is $0.57\ \mu$m, which is 43 times greater than the width $l_\mathrm{0}$ of tensile domain walls. The period of the striped network  is $0.21\ \mu$m, which is 17 times greater than the width of domain walls with $\beta=\pm\pi/6$. Under such conditions, the two-chain Frenkel-Kontorova model should still be adequate and we can expect that the accuracy of our estimates of the critical elongations is about 20\% and mostly comes from the uncertainty in the barrier $V_\mathrm{max}$ to relative sliding of the layers
\cite{Lebedeva2019, Lebedeva2019a}.

Since the triangular and striped phases have different symmetries, the transition between them is of the first order. Thus, they can coexist in the same sample, as indeed observed in the experiments \cite{Alden2013,Kisslinger2015}. In the phase boundary, domain walls aligned in two different zigzag directions in the triangular phase gradually become parallel and aligned in the same armchair direction, while the separation between the walls in the third zigzag direction increases and such walls finally disappear.
The relative area of bilayer beloning to domain walls  increases in the transition by 48\%. The ratio of critical elongations $\epsilon_\mathrm{c1}/\epsilon_\mathrm{c0}$ is determined only by Poisson's ratio. Therefore, the phase transion between the regular triangular and striped incommensurate phases can be expected also for other 2D materials with the shape of the surface of the interlayer interaction energy described by Eq. (\ref{eq_pes}).

Let us now discuss the limit of large elongations when the interlayer interaction does not perturb significantly the structure of the layers \cite{Lebedev2017}. In this limit, the system in the regular triangular incommensurate phase should gradually approach the fully incommensurate state in which the bottom layer is uniformly stretched and the upper layer is relaxed. The period of such a superstructure is $L_0=l\sqrt{3}/\epsilon$  (Fig. \ref{fig:length}). The elastic energy gain with respect to the commensurate bilayer is $\Delta W_\mathrm{el} =-k\epsilon^2/(1-\nu)$. The interlayer interaction energy in this system corresponds to the average over the potential energy surface (\ref{eq_pes}), $\Delta W_\mathrm{int}=V_\mathrm{in}= 3 V_\mathrm{max}$. Therefore, the relative energy of the fully incommensurate bilayer is $\Delta W_\mathrm{0} =\Delta W_\mathrm{el} + \Delta W_\mathrm{int} = -k\epsilon^2/(1-\nu) + 3 V_\mathrm{max}$.

As a limit for the striped incommensurate phase, we should consider the state in which the bottom layer is uniformly stretched and the upper layer is displaced along the minimum energy paths between adjacent minima of the potential energy surface (\ref{eq_pes}) in such a way that the displacement across the stripes scales linearly with the distance. The interlayer interaction energy in this case is given by the average over the minimum energy path $\Delta W^\mathrm{s}_\mathrm{int} =V_\mathrm{av} =  V_\mathrm{max}(3-9\sqrt{3}/(2\pi))$. If the displacement between the equivalent AB energy minima occurs at the distance $L$ across the stripes, the tensile strain in the upper layer in this direction  equals  $\epsilon-\epsilon_\mathrm{u}$, where $\epsilon_\mathrm{u}=l\sqrt{3}/L$, and the shear strain $\tau_\mathrm{u}=l/L$. Therefore, the elastic energy compared to the commensurate bilayer is given by 
\begin{equation} \label{eq_Wels2}
\begin{split}
 &\Delta W^\mathrm{s}_\mathrm{el} =-\frac{k\epsilon\epsilon_u}{1-\nu}+\frac{k\epsilon_\mathrm{u}^2}{2(1-\nu^2)} +\frac{k\tau_\mathrm{u}^2}{4(1+\nu)}\\&= -\frac{\sqrt{3} k\epsilon}{1-\nu}\frac{l}{L}+\frac{(7-\nu)k}{4(1-\nu^2)}\left(\frac{l}{L}\right)^2.
\end{split}
\end{equation}
The minimal elastic energy 
\begin{equation} \label{eq_Ws0_3}
\Delta W^\mathrm{s}_\mathrm{el,0}=-\frac{3(1+\nu)}{(1-\nu)(7-\nu)}k\epsilon^2
\end{equation}
is reached for the superstructure period (Fig. \ref{fig:length})
\begin{equation} \label{eq_Ws0_3}
L^\mathrm{s}_0=\frac{7-\nu}{2\sqrt{3}(1+\nu)}\frac{l}{\epsilon}.
\end{equation}  

It is clear that the triangular incommensurate phase becomes energetically favourable over the striped one at the elongation
\begin{equation} \label{eq_eps2}
\begin{split}
\epsilon_\mathrm{c2}=\frac{3}{2}\sqrt{\frac{\sqrt{3}(7-\nu)V_\mathrm{max}}{2\pi k}}=1.11\cdot10^{-2}.
\end{split}
\end{equation}
Therefore, it should be expected that the reentrant phase transition takes place in bilayer graphene with one stretched layer at the relative biaxial elongation $\sim10^{-2}$  (Fig. \ref{fig:length}).

In summary, we predict that upon biaxial stretching of one of the layers of bilayer graphene, the commensurate-incommensurate phase transition to the phase with the regular triangular domain wall network is followed by the transition to the phase with the striped domain wall network and then by the reentrant transition to the triangular phase. Recently the local change of the domain wall shape using the scanning tunneling \cite{Jiang2018} and 
atomic force microscope \cite{Yankowitz2014} tips has been achieved. The phase transitions proposed here modify the whole structure of the domain wall network and thus can be used to manipulate electronic, optical and magnetic properties of bilayer graphene promising for application in graphene-based nanoelectronics.

\begin{acknowledgments}
AMP acknowledges the Russian Foundation for Basic Research (Grant 18-02-00985).
\end{acknowledgments}

\bibliography{rsc}
\end{document}